\documentclass[twocolumn,english]{revtex4-2}
\usepackage[T1]{fontenc}
\usepackage[latin9]{inputenc}
\usepackage{float}
\usepackage{amsmath}
\usepackage{amssymb}
\usepackage{graphicx}

\makeatletter

\providecommand{\tabularnewline}{\\}

\makeatother

\usepackage{babel}
\begin{document}
\title{A semi-empirical analysis of the paramagnetic susceptibility of solid
state magnetic clusters}
\author{\textsc{Lauro B. Braz, Fernando A. Garcia}}
\affiliation{$^{1}$Instituto de Física da Universidade de São Paulo (IFUSP), Universidade
de São Paulo, São Paulo-SP, 05508-090, Brazil.}
\begin{abstract}
Recent developments in the synthesis of new magnetic materials lead
to the discovery of new quantum paramagnets. Many of these materials,
such as the perovskites Ba$_{4}$LnMn$_{4}$O$_{12}$ (Ln = Sc or
Nb), Ba$_{3}$Mn$_{2}$O$_{8}$, and Sr$_{3}$Cr$_{2}$O$_{8}$ present
isolated magnetic clusters with strong intracluster interactions but
weak intercluster interactions, which delays the onset of order to
lower temperatures ($T$). This offset between the local energy scale
and the magnetic ordering temperature is the hallmark of magnetic
frustration. At sufficient high-$T$, the paramagnetic susceptibility
($\chi$) of frustrated cluster magnets can be fit to a Curie-Weiss
law, but the derived microscopic parameters cannot in general be reconciled
with those obtained from other methods. In this work, we present an
analytical microscopic theory to obtain $\chi$ of dimer and trimer
cluster magnets, the two most commonly found in literature, making
use of suitable Heisenberg-type Hamiltonians. We also add intercluster
interactions in a mean-field level, thus obtaining an expression to
the critical temperature of the system and defining a new effective
frustration parameter $f_{\text{eff}}$. Our method is exemplified
by treating the $\chi$ data of some selected materials. 
\end{abstract}
\maketitle

\section{Introduction}

The magnetism of solids encompass a really broad research field, ranging
from studies of magnets for applications to the investigation of the
fundamentals of electronic interactions in matter \citep{stohrMagnetismFundamentalsNanoscale2006,whiteQuantumTheoryMagnetism2007}.
Energy scales of the magnetic interactions are set by the exchange
constants $J$, which are mainly determined by the nature of the interacting
spins and the electronic structure of the magnetic active atom coordination
structure \citep{goodenoughMagnetismChemicalBond1963}. 

To estimate $J$ is an important step towards understanding the magnetism
of a particular material. The first approach to this problem is given
by the Curie-Weiss analysis of the material paramagnetic susceptibility
$\chi$. For the vast majority of magnetic solid state materials,
$\chi$ as a function of temperature ($T$) is fairly described by
the Curie-Weiss expression (Equation \ref{eq:curie})

\begin{equation}
\chi(T)=\frac{C}{T-\theta_{\text{CW}}}\label{eq:curie}
\end{equation}

where $C$ and $\theta_{\text{CW}}$ are the Curie and Curie-Weiss
constants, respectively. As is well known, $\theta_{\text{CW}}$ can
be connected to $J$ and, in a mean field approach, to the magnetic
ordering temperature of solids \citep{whiteQuantumTheoryMagnetism2007},
whereas the value of $C$ relates to the single ion spin configuration
in the solid. Exceptions, however, do exist for which the Curie-Weiss
approach cannot provide a physically meaningful set of parameters.
Recently, the magnetic properties of a large class of perovskite-type
materials were reviewed \citep{nguyenHexagonalPerovskitesQuantum2021}
providing many important examples where the set of the obtained $C$
and $\theta_{\text{CW}}$ parameters do not connect well with the
known properties of the materials. This happens in respective of the
apparent good fittings of the high-$T$ $\chi$ data to the Curie-Weiss
expression (Eq. \ref{eq:curie}). Illustrative examples are provided
by Ba$_{4}$$Ln$Mn$_{3}$O$_{12}$ ($Ln=$ Sc or Nb) $\chi$ data
\citep{yinNew10HPerovskites2017c,nguyenTrimersMnO6Octahedra2019c},
for which the obtained $C$ values are much too low to be compared
with the expectation of $S=2$ spins from Mn$^{3+}$ cations. If the
obtained $C$ values are not reliable, it also raises questions about
the obtained $\theta_{\text{CW}}$ parameters. 

The Ba$_{4}$$Ln$Mn$_{3}$O$_{12}$ ($Ln=$ Sc or Nb) materials shared
the same crystal structure (see Figure \ref{fig:full}) which host
as building blocks isolated clusters of magnetic cations, sitting
within face sharing Oxygen octahedra. This building block displays
a large number of exchange paths which in turn contribute to large
$J$'s \citep{goodenoughMagnetismChemicalBond1963}. The local (within
the cluster) magnetic interactions are strong but the interclusters
interactions are weak, since the clusters are well separated in the
structure. The onset of order, if observed, is thus delayed to low
temperatures. This is the hallmark of frustrated magnetism \citep{moessnerGeometricalFrustration2006}.

A feature of frustrated magnetism is that if the frustration is strong
enough, it may compete with quantum fluctuations for the system ground
state, giving rise to exotic types of quantum magnetism, such as quantum
spin liquids \citep{balentsSpinLiquidsFrustrated2010,zhouQuantumSpinLiquid2017}
or Bose-Einstein condensates of spin excitations \citep{nikuniBoseEinsteinCondensationDilute2000,aczelFieldInducedBoseEinsteinCondensation2009}.
Thus, materials hosting this type of building blocks are good platforms
to search for exotic magnetism as noted in Ref. \citep{nguyenHexagonalPerovskitesQuantum2021}.
The usual way to quantify frustration is by determining the frustration
parameter $f$, or a lower bound to $f$, which requires a trustworthy
determination of the system's energy scales.

In this work, we propose to analyze the paramagnetic data of materials
hosting magnetic clusters adopting Heisenberg-type Hamiltonians \citep{whiteQuantumTheoryMagnetism2007}
parametrized by exchange constants $J_{n}$ , with the intercluster
interactions taken into account by a mean field approach. This is
an alternative, semi-empirical approach which, while already tested
in the case of spin dimers \citep{deisenhoferStructuralMagneticDimers2006,singhSingletGroundState2007},
is not particularly explored in the case of spin trimers \citep{yinNew10HPerovskites2017c,nguyenTrimersMnO6Octahedra2019c}.

In section \ref{sec:models}, we provide a general perspective on
the proposed methodology and then we illustrate our approach by fitting
the $\chi$ data of a series of materials hosting magnetic clusters
in section \ref{sec: results}. Finally, we move to discuss and summarize
our results. Most interesting, we define a new frustration parameter
$f_{\text{eff}}$ in terms of the energy scale of the effective intercluster
interactions. In doing so, we show some frustrated magnets are unlikely
to display quantum many body states at further lower temperatures,
while other remain as strong candidates to host this type of Physics. 

\section{Models}

\label{sec:models} 
\begin{figure*}
\centering{}\includegraphics[width=1\textwidth]{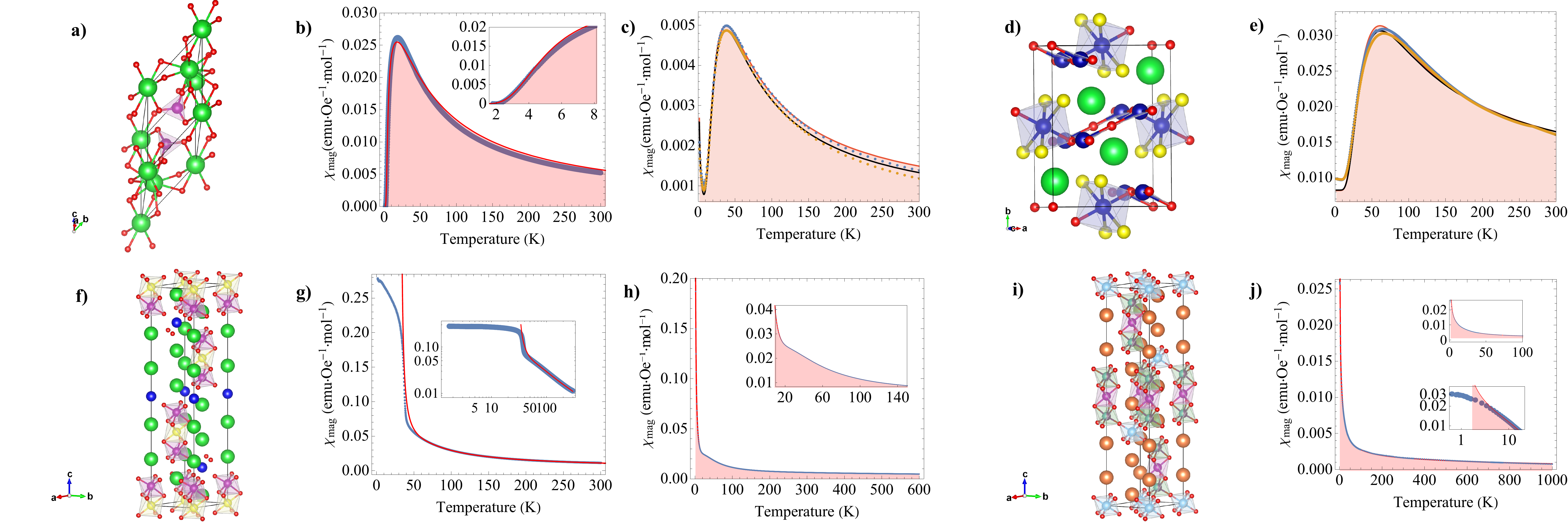}
\caption{$(a)$ Structural model of the Ba$_{3}$Mn$_{2}$O$_{8}$ and Sr$_{3}$Cr$_{2}$O$_{8}$
materials (symmetry group R$\bar{3}$m). The spheres represent Mn/Cr
(purple; magnetic), Ba (green), and O (red) atoms. $(b)$ Ba$_{3}$Mn$_{2}$O$_{8}$
$\chi$ data (blue dots) and theoretical fit (red line) $(c)$ Sr$_{3}$Cr$_{2}$O$_{8}$
$\chi$ data with blue as yellow dots representing, respectively,
data obtained with the field applied parallel and perpendicular to
the sample's $ab$ plane. The red and black lines represent the respective
theoretical fittings. $(d)$ Sr$_{3}$Co$_{3}$S$_{2}$O$_{3}$ structural
model. The spheres represent Co (blue; magnetic), Sr (green), O (red),
and S (yellow) atoms. $(e)$ As in $(c)$ but for the Sr$_{3}$Co$_{3}$S$_{2}$O$_{3}$
material. $(f)$ Structural model of the Ba$_{4}$ScMn$_{3}$O$_{12}$
and Ba$_{4}$NbMn$_{3}$O$_{12}$ materials (symmetry group R$3$m).
The spheres represent Mn (purple/yellow; magnetic), Ba (green), O
(red), and Nb/Sc (blue) atoms. $(g)-(h)$ Susceptibility data (blue
dots) and respective fitting (red line) for the Ba$_{4}$ScMn$_{3}$O$_{12}$,
and Ba$_{4}$NbMn$_{3}$O$_{12}$ materials, respectively. $(i)$
Structural model of BaTi$_{0.5}$Mn$_{0.5}$O$_{3}$. The spheres
represent Mn (purple; magnetic), Ti (light blue), O (red), and Ba
(orange) atoms. $(j)$ Susceptibility data (blue dots) and fittings
(red line) for BaTi$_{0.5}$Mn$_{0.5}$O$_{3}$. \label{fig:full}}
\end{figure*}

Our approach is mainly intended to the description
of the magnetic properties of solids hosting magnetic clusters as
exemplified by the crystal structures in Figure \ref{fig:full}$(a)\text{, }(d)\text{, }(f)\text{ and }(i)$.
In all cases, the magnetic cations are connected by multiple exchange
paths, causing the local magnetic interaction to be strong. We thus
treat single cluster magnetic properties by means of a microscopic
Heisenberg-type Hamiltonian:

\begin{equation}
\mathcal{H}=\sum_{ij}J_{ij}\boldsymbol{S}_{i}\cdot\boldsymbol{S}_{j}\label{eq:heisemberg}
\end{equation}

where the indexes $i$ and $j$ run within the cluster sites, $\boldsymbol{S}_{i}$
are the spin operators of the magnetic cations and $J_{ij}$ is the
exchange constant between spins $i$ and $j$. For instance, applied
to the case of the magnetic trimers, Equation \ref{eq:heisemberg}
reads:

\begin{equation}
\mathcal{H}_{\text{trim}}=J_{12}S_{1}\cdot S_{2}+J_{23}S_{2}\cdot S_{3}+J_{13}S_{1}\cdot S_{3}.\label{eq:hamtrimer}
\end{equation}

In this work, we shall also illustrate our approach with magnetic
dimers, that can be described by:

\begin{equation}
\mathcal{H}_{\text{dim}}=J_{1}S_{1}\cdot S_{2}.\label{eq:hamdimer}
\end{equation}

Once the proper Hamiltonian is determined, we consider a weak magnetic
field $H$ and evaluate the system energy levels up to first order
in the field:

\begin{equation}
E_{n}=E_{n}^{(0)}+HE_{n}^{(1)}+\mathcal{O}(2)\label{eq:energyorderfield}
\end{equation}

and the results are applied to calculate the system magnetic bare
susceptibility in the usual units of $\text{emu}\cdot\text{mol}^{-1}\cdot\text{Oe}^{-1}$.

\begin{equation}
\bar{\chi}(\beta)=N_{A}k_{B}\mu_{B}^{2}\frac{\sum_{n}e^{-E_{n}^{(0)}\beta}\left(E_{n}^{(1)}/\mu_{B}\right)^{2}}{\sum_{n}e^{-E_{n}^{(0)}\beta}}\beta\label{eq:susceptibility}
\end{equation}

where $N_{A}$ is the Avogadro number, $k_{\text{B}}$ is the Boltzmann
constant, $\mu_{\text{B}}$ is the Bohr magneton, $\beta=1/k_{B}T$
and $E_{n}^{(0)}$and $E_{n}^{(1)}$ are, respectively, the zeroth
and first order eigenvalues of $\mathcal{H}$ (Equation \ref{eq:heisemberg})
as defined by Equation \ref{eq:energyorderfield}. In many situations
of interest, Equations \ref{eq:hamtrimer} and \ref{eq:hamdimer}
can be diagonalized analytically (see appendix \ref{sec:appendixA})
and thus the perturbation theory to write down $\bar{\chi}$ can be
carried out straightforwardly.

The expression obtained from Equation \ref{eq:susceptibility} contains
the $J_{n}$ constants as free parameters. The next step is to treat
the cluster-cluster interaction in a mean field approximation:

\begin{equation}
\chi=\frac{\bar{\chi}}{1-\lambda\bar{\chi}}\label{eq:rpchi}
\end{equation}

where $\lambda$ is the molecular field parameter, describing the
inter-cluster interaction. As we shall discuss, in many instances
it is $\lambda$ which should be adopted to estimate the values of
the frustration parameter $f$.

\section{Results}

\label{sec: results} Now that we have developed the appropriate tools
to model clusters, the next step is to apply the theory to better
understand some model systems. We start taking spin dimers into account,
as a way to illustrate our approach in an already familiar environment,
and then we move to investigate spin trimers in some selected hexagonal
perovskites \citep{nguyenHexagonalPerovskitesQuantum2021}. These
low symmetry systems are prone to exhibit rotated Oxygen octahedra
coordinating transition metal cations, which will precisely give rise
to the multiplicity of the exchange paths that cause the formation
of strongly coupled magnetic clusters. We end this section analyzing
a system exhibiting both spin dimers and spin trimers.

\subsection{Dimers}

\label{sec:expdimer} We select the Ba$_{3}$Mn$_{2}$O$_{8}$ \citep{uchidaSingletGroundState2001,uchidaHighfieldMagnetizationProcess2002,tsujiiSpecificHeatSpindimer2005},
Sr$_{3}$Cr$_{2}$O$_{8}$ \citep{aczelFieldInducedBoseEinsteinCondensation2009},
and Sr$_{3}$Co$_{3}$S$_{2}$O$_{3}$ \citep{laiCoexistenceSpinOrdering2017f}
spin dimer materials to our analysis. The same trigonal structure,
described by the space group R$\bar{3}$m, are shared by the two former
materials and is depicted in Figure \ref{fig:full}$(a)$. The Sr$_{3}$Co$_{3}$S$_{2}$O$_{3}$
structure is described by the space group Pbam and is presented in
Figure \ref{fig:full}$(d)$.

Focusing on the trigonal materials, charge balance in the systems
dictate that we have Mn$^{5+}$and Cr$^{5+}$ cations which carry,
respectively, $S=1$ and $S=1/2$ spins. The Ba$_{3}$Mn$_{2}$O$_{8}$
total $\chi$ is modeled by adopting one exchange constant $J_{1}$
and one $\lambda$ molecular field parameter plus a constant diamagnetic
contribution $\chi_{0}$ as in Equation \ref{eq:suscdim}:

\begin{equation}
\chi(T)=q_{\text{dim}}\chi_{\text{dim}}+\chi_{0}\label{eq:suscdim}
\end{equation}
where $q_{\text{dim}}$ is the molar fraction of the dimer specimen
in the sample and $\chi_{\text{dim}}$ is obtained from \ref{eq:rpchi}
(with $\bar{\chi}$ as suitable to case of dimers, see Equation \ref{eq:endim})
and carries the $J_{1}$ and $\lambda$ parameters. The fitting and
experimental data are presented in Figure \ref{fig:full}$(b)$ and
the obtained parameters are $J_{1}/k_{B}=17.66(7)$ K and $\lambda=-16.35(9)$
Oe$\cdot$mol$\cdot$emu$^{-1}$.

In the case of the Sr$_{3}$Cr$_{2}$O$_{8}$ material \citep{aczelFieldInducedBoseEinsteinCondensation2009},
we model to Equation \ref{eq:suscdim} an orphan spin contribution
($C/T$) to account for the observed rise in $\chi$ at low temperatures.
The fitting and experimental data are presented in Figure \ref{fig:full}$(c)$
and the obtained parameters are $J_{1}/k_{B}=61.5(2)$ K, $\lambda=-5(1)$
Oe$\cdot$mol$\cdot$emu$^{-1}$, $C=4.06(9)\cdot10^{-3}$ emu$\cdot$K/mol,
and $\chi_{0}=1.9(1)\cdot10^{-4}$ emu/mol Cr.

Lastly, as discussed in Ref. \citep{laiCoexistenceSpinOrdering2017f},
the Sr$_{3}$Co$_{3}$S$_{2}$O$_{3}$ compound was expected to hold
an effective spin $S=3/2$ because of the Co$^{2+}$ valence, which
is again deduced from charge balance. Due to the octahedral crystal
field, however, Co$^{2+}$ cations carry low $S=1/2$ spins. The total
$\chi$ is modeled by Equation \ref{eq:suscdim} and the obtained
parameters for the FC (ZFC) susceptibility data are $J_{1}/k_{B}=97.9(2)$
K ($J_{1}/k_{B}=99.6(2)$ K), $\lambda=2.50(16)$ Oe$\cdot$mol$\cdot$emu$^{-1}$
($\lambda=0.94(11)$ Oe$\cdot$mol$\cdot$emu$^{-1}$), and $\chi_{0}=0.00829(3)$
emu/mol Co ($\chi_{0}=0.00824(3)$ emu/mol Co).Slightly distinct values
are reported in Ref. \citep{laiCoexistenceSpinOrdering2017f} because
here we included the intercluster interactions, which we found to
be ferromagnetic ($\lambda>0$). The fitting and experimental data
are presented in Figure \ref{fig:full}$(e)$.

\subsection{Trimers}

\label{sec:exptrimer} Two materials presenting magnetic trimers were
selected: Ba$_{4}$ScMn$_{3}$O$_{12}$ and Ba$_{4}$NbMn$_{3}$O$_{12}$.
The materials share the same crystal structure which is depicted in
Figure \ref{fig:full}$(f)$. In both cases, we assume the presence
of Mn$^{4+}$cations carrying $S=3/2$ spins and Hamiltonian \ref{eq:hamtrimer}
is adopted to describe the trimer magnetism. It should be clear, however,
that we can adopt $J_{12}=J_{23}=J_{1}$ and $J_{13}=J_{2}=\alpha J_{1}$,
where $0\leq\alpha\leq1$. This assumption make it possible to perform
an exact diagonalization of \ref{eq:hamtrimer}. In principle, one
could think of $\alpha$ as a small number, but the large multiplicity
of exchange paths cause $\alpha$ to be $>0.5$ for magnetic cations
within face sharing octahedra. Intertrimer interactions are described
by just one molecular field parameter $\lambda$. We adopt the model
susceptibility:

\begin{equation}
\chi(T)=q_{\text{trim}}\chi_{\text{trim}}+\chi_{0}\label{eq:susctrim}
\end{equation}
where $q_{\text{trim}}$ is the molar fraction of the trimer specimen
in the sample and $\chi_{\text{trim}}$ is obtained from equation
\ref{eq:susctrim} (see Equation \ref{eq:entrim} for more details).
One should note that $\chi_{\text{trim}}$ carries three parameters:
the exchange constants $J_{1}$ and $J_{2}$ and $\lambda$.

In the case of Ba$_{4}$ScMn$_{3}$O$_{12}$, the following parameters
are obtained: $J_{1}/k_{B}=199.44(2)$ K, $\alpha=0.60041(1)$, $\chi_{0}=-2.67(4)\cdot10^{-5}$
emu$\cdot$K/mol/Oe, and a negligible value of $\lambda$. As for
Ba$_{4}$NbMn$_{3}$O$_{12}$ , we obtain: $J_{1}/k_{B}=172(1)\text{ K}$,
$\alpha=0.7651(1)$, $\lambda=77.4(5)$ mol$\cdot$Oe/emu/K, and $\chi_{0}=-0.00263(3)$
emu$\cdot$K/mol/Oe. The fittings results are compared to the data
in Figure \ref{fig:full}$(f)-(g)$, respectively. The significant
value of $\lambda$ found in the later case is in agreement with the
observed critical temperature of $\approx32$ K determined for Ba$_{4}$NbMn$_{3}$O$_{12}$.

\subsection{Dimers and trimers}

\label{sec:mixed} We now turn to the challenging case offered by
the BaTi$_{0.5}$Mn$_{0.5}$O$_{3}$ \citep{garciaMagneticDimersTrimers2015f,cantarinoDynamicMagnetismDisordered2019}
material, which presents spin dimers, trimers, and orphans. Its crystal
structure is depicted in Figure \ref{fig:full}$(i)$. The proposed
model to $\chi$ assumes the following form:

\begin{equation}
\chi(T)=q_{\text{trim}}\chi_{\text{trim}}+q_{\text{dim}}\chi_{\text{dim}}+q_{\text{orp}}\chi_{\text{orp}}+\chi_{0}\label{eq:suscmix}
\end{equation}

where $q_{\text{orp}}$ is the molar fraction of the orphan spins
in the sample and $\chi_{\text{orp}}$ is the orphan spin susceptibility.
As discussed elsewhere \citep{garciaMagneticDimersTrimers2015f,cantarinoDynamicMagnetismDisordered2019},
the fractions are statistically determined to be: $q_{\text{trim}}=q_{\text{orp}}=1/16$,
and $q_{\text{dim}}=2/16$. We adopt a Curie-Weiss form to $\chi_{\text{orp }}$
(with $C_{\text{orp}}$ and $\theta_{\text{orp}}$ parameters). The
obtained fitting is compared to the data in Figure \ref{fig:full}$(j)$
and the obtained cluster parameters are: $J_{1}/k_{B}=201(2)$ K,
$\alpha=0.85(2)$, and $\lambda=48(15)$ mol$\cdot$Oe/emu/K (calculated
considering only timer-trimer interactions), while the orphan spin
parameters are: $\theta_{\text{orp}}=14.4(5)$ K and $C_{\text{orp}}=0.217(2)$
emu$\cdot$K/mol/Oe.

\begin{table*}
\caption{Summary of experimentally obtained parameters within our model paramagnetic
susceptibility. \label{tbl:summary}}

\centering{}%
\begin{tabular*}{1\textwidth}{@{\extracolsep{\fill}}cccccccc}
\multicolumn{7}{c}{} & \tabularnewline
\hline 
Material & $J_{1}/k_{B}$ (K) & $\alpha$ ($J_{2}/J_{1}$) & $\lambda$ (Oe$\cdot$mol/emu) & $s_{\text{eff}}$ & $T_{\text{C,eff}}$ (K) & $f_{\text{eff}}$ & $f_{\text{CW}}$\tabularnewline
\hline 
Ba$_{3}$Mn$_{2}$O$_{8}$ & $17.66(7)$ & - & $-16.35(9)$ & $1$ & $16.36(9)$ & $9.14$ & $18.7$\tabularnewline
Sr$_{3}$Cr$_{2}$O$_{8}$ & $61.5(2)$ & - & $-5(1)$ & $1/2$ & $1.7(4)$ & $0.90$ & $26.2$\tabularnewline
Sr$_{3}$Co$_{3}$S$_{2}$O$_{3}$ & $99.6(2)$ & - & $1.82(17)$ & $1/2$ & $2.4(2)$ & $0.63$ & $42.8$\tabularnewline
Ba$_{4}$ScMn$_{3}$O$_{12}$ & $199.44(2)$ & $0.60041(1)$ & $-0.05543(5)$ & $1/2$ & $0.0208(2)$ & $0.01$ & $0.41$\tabularnewline
Ba$_{4}$NbMn$_{3}$O$_{12}$ & $172(1)$ & $0.7651(1)$ & $77.4(5)$ & $1/2$ & $29.0(1)$ & $0.91$ & $0.02$\tabularnewline
BaTi$_{0.5}$Mn$_{0.5}$O$_{3}$ & $201(2)$ & $0.85(2)$ & $48(15)$ & $1/2$ & $18(6)$ & $180$ & $1024$\tabularnewline
\hline 
\hline 
 & \multicolumn{2}{c}{} & \multicolumn{2}{c}{} & \multicolumn{2}{c}{} & \tabularnewline
\end{tabular*}
\end{table*}

\section{Discussion}

We would like to discuss two aspects or our results: $i)$ the set
of obtained parameters and $ii)$ the interpretation of $\lambda$
and its relation to the frustration parameter $f$, which is of great
relevance to the characterization of putative quantum phases associated
to these materials. We start with the later. 

By comparing the high temperature ($\beta\rightarrow0$) limit of
Equation \ref{eq:rpchi} to the the Curie-Weiss law (Equation \ref{eq:curie}),
one reaches the expression $\bar{\theta}_{\text{CW}}=\lambda C$,
which in turn can be applied to define an effective exchange constant
$J_{\text{eff }}$for the intercluster interaction as $\lambda\equiv\frac{J_{\text{eff}}k_{B}}{g^{2}\mu_{B}^{2}N_{A}}$.
This expression gives a well defined meaning to $\lambda$ as the
parameter describing long range intercluster interactions.

Indeed, in forming the clusters, most of the magnetic degrees of freedom
are frozen at low temperatures and the remaining degrees of freedom
interact with an energy scale characterized by $\bar{\theta}_{\text{CW}}$.
Therefore, the correct estimate to the system frustration should be
given by a new effective frustration parameter $f_{\text{eff}}\equiv|\frac{\bar{\theta}_{\text{CW}}}{T_{\text{L}}}|$
where $T_{\text{L}}$ is the lowest temperature at which the system
is still in the paramagnetic state. If order is not observed, $T_{\text{L}}$
is given by the lowest experimentally achieved temperature and the
$f_{\text{eff}}$ parameter thus obtained is a lower bond to $f_{\text{eff}}$.
Moreover, we have now a way to estimate the temperature for long range
magnetic order that will be triggered by the intercluster interactions.
We shall name this the effective critical temperature, $T_{\text{C,eff}}$.
Adopting the overall mean field relation $|\theta_{\text{CW}}|=T_{\text{C }}$,
where $T_{\text{C }}$ is the magnetic critical temperature, we write
$|\bar{\theta}_{\text{CW}}|=T_{\text{C,eff}}$ and then $|\bar{\theta}_{\text{CW}}|=T_{\text{C,eff}}=\lambda C$.
To obtain $C$, we consider that the remaining degrees of freedom
of the magnetic clusters should be associated with an effective spin
$s_{\text{eff}}$ , thus we can write:

\begin{equation}
T_{\text{C,eff}}=|J_{\text{eff}}|\frac{s_{\text{eff}}(s_{\text{eff}}+1)}{3}\label{eq:Tc}
\end{equation}

We propose that $s_{\text{eff}}$ should be calculated as either the
expectation value of the $S_{z}$ operator of the cluster ground state
spin state in the case of trimers, or by the first excited total spin
state of the cluster, in the case of dimers. Concerning the later,
the necessity of considering the first excited state in the case of
dimers can be understood as follows: the first order energy correction
for the spin excitation in a given magnetic field $H$ is of the type
$E_{0}^{(1)}=-g\mu_{B}Hs_{\text{eff}}$. For an antiferromagnetic
dimer, $s_{\text{eff}}$ is always $0$ if one adopts the spin ground
state to calculate $s_{\text{eff}}$. Our assumption about $s_{\text{eff}}$
thus means that the system magnetism at low-$T$ is dominated only
by the ground state spin configuration except when it turns out be
null, when one should peak the first excited state. In the appendix
\ref{sec:appendixB} we show how to determine $s_{\text{eff}}$ for
trimers. 

All obtained parameters are shown in table \ref{tbl:summary}. We
also list $f_{\text{CW}}$ parameters which are the frustration parameters
obtained from the usual Curie-Weiss analysis. The values of $J_{n}$
clearly distinguish between the exchange constants due to magnetic
interactions mediated by corner sharing octahedra, the case of the
selected dimer based materials, and face sharing octahedra, the case
of selected trimer based materials. Intercluster interactions of both
FM and AFM types are present, but are definitely small energy scales
when compared to the intracluster interactions. Notwithstanding, these
are the energy scales that will control the onset of magnetic order
at sufficient low temperatures.

Comparing $f_{\text{eff}}$ and $f_{\text{CW}}$, one can conclude
that all dimer based materials are at most moderately frustrated magnetic
systems (the case of Ba$_{3}$Mn$_{2}$O$_{8}$) if the intercluster
energy scale is considered. It is therefore unlikely that a quantum
magnetic state will emerge from these materials. Interesting glassy
behavior, however, could be expected.

The case of the trimer based materials is even more revealing. Ba$_{4}$NbMn$_{3}$O$_{12}$
order at $T\approx32$ K and is not a frustrated material. Ba$_{4}$ScMn$_{3}$O$_{12}$,
on the other hand, is expected to order only about $T\approx0.02$
K, a temperature at which quantum effects could tune the system into
an exotic ground state. The intercluster interactions, however, are
rather weak making it unlikely. The mixed dimer/trimer material BaTi$_{0.5}$Mn$_{0.5}$O$_{3}$
remain characterized as a strongly frustrated magnet with relatively
large intercluster interactions. As already observed \citep{cantarinoDynamicMagnetismDisordered2019},
the system disorder is large, which may hinder the appearance of a
quantum spin liquid, although the physics will be that of a correlated
disordered quantum magnet. Thus, further investigations at low $T$
are invited.

\section{Conclusion and perspectives}

\label{sec:conclusion} The Curie-Weiss law has great use to analyze
the magnetism of very general systems. It fails, however, in giving
a microscopic view of the problem. We found the later to be particularly
critical to determine the energy scale of the magnetic interaction
of materials hosting magnetic clusters. Therefore, in this work we
proposed a methodology to study the susceptibility of cluster magnets
which considers intracluster and intercluster interactions. 

Our methods were applied to a series of materials and the obtained
parameters were discussed and put into perspective. Our semiempirical
approach to the problem could identify potential material candidates
to low temperature explorations in the search for quantum many body
ground states. Our main proposal is that intercluster interactions
must be taken into account while discussing the level of frustration
in the system. This is the energy scale that correctly sets the odds
for a system to display a quantum many body state.

In particular, the trimmer system Ba$_{4}$ScMn$_{3}$O$_{12}$, which
present a very small inter-cluster interaction, with a predicted ordering
temperature of $\approx0.02$ K, is unlikely to display this type
of physics, but the case of mixed dimer/trimmer system BaTi$_{0.5}$Mn$_{0.5}$O$_{3}$
remain open for further experimental investigation. 

Lastly, one could argue that our model, in comparison to Curie-Weiss
analysis, is only adding fitting parameters. We point out, however,
that our approach starts from a microscopic model and then scales
to the description of macroscopic susceptibility data. The adopted
parameter set are thus not arbitrary. In fact, the present method
proposes a more meaningful analysis of magnetic susceptibility data
of systems presenting magnetic clusters.

\section{Acknowledgments}

We thank H. Tanaka and collaborators, A. Aczel and collaborators \citet{aczelFieldInducedBoseEinsteinCondensation2009},
K. To Lai and M. Valldor \citet{laiCoexistenceSpinOrdering2017f}
and E. Komleva and collaborators \citet{komlevaThreesiteTransitionmetalClusters2020}
who provided us with the Ba$_{3}$Mn$_{2}$O$_{8}$, Sr$_{3}$Cr$_{2}$O$_{8}$,
Sr$_{3}$Co$_{3}$S$_{2}$O$_{8}$, and Ba$_{4}$NbMn$_{3}$O$_{12}$
data, respectively. The financial support from Fundação de Amparo
a Pesquisa do Estado de São Paulo is acknowledged by L.B.B. (Grant
No. 2019/27555-9) and F.A.G. (Grant No. 2019/25665-1).

\bibliographystyle{apsrev4-2}
\bibliography{2021Braz_semiempirical_magclusters}

\begin{thebibliography}{20}%
\makeatletter
\providecommand \@ifxundefined [1]{%
 \@ifx{#1\undefined}
}%
\providecommand \@ifnum [1]{%
 \ifnum #1\expandafter \@firstoftwo
 \else \expandafter \@secondoftwo
 \fi
}%
\providecommand \@ifx [1]{%
 \ifx #1\expandafter \@firstoftwo
 \else \expandafter \@secondoftwo
 \fi
}%
\providecommand \natexlab [1]{#1}%
\providecommand \enquote  [1]{``#1''}%
\providecommand \bibnamefont  [1]{#1}%
\providecommand \bibfnamefont [1]{#1}%
\providecommand \citenamefont [1]{#1}%
\providecommand \href@noop [0]{\@secondoftwo}%
\providecommand \href [0]{\begingroup \@sanitize@url \@href}%
\providecommand \@href[1]{\@@startlink{#1}\@@href}%
\providecommand \@@href[1]{\endgroup#1\@@endlink}%
\providecommand \@sanitize@url [0]{\catcode `\\12\catcode `\$12\catcode
  `\&12\catcode `\#12\catcode `\^12\catcode `\_12\catcode `\%12\relax}%
\providecommand \@@startlink[1]{}%
\providecommand \@@endlink[0]{}%
\providecommand \url  [0]{\begingroup\@sanitize@url \@url }%
\providecommand \@url [1]{\endgroup\@href {#1}{\urlprefix }}%
\providecommand \urlprefix  [0]{URL }%
\providecommand \Eprint [0]{\href }%
\providecommand \doibase [0]{https://doi.org/}%
\providecommand \selectlanguage [0]{\@gobble}%
\providecommand \bibinfo  [0]{\@secondoftwo}%
\providecommand \bibfield  [0]{\@secondoftwo}%
\providecommand \translation [1]{[#1]}%
\providecommand \BibitemOpen [0]{}%
\providecommand \bibitemStop [0]{}%
\providecommand \bibitemNoStop [0]{.\EOS\space}%
\providecommand \EOS [0]{\spacefactor3000\relax}%
\providecommand \BibitemShut  [1]{\csname bibitem#1\endcsname}%
\let\auto@bib@innerbib\@empty
\bibitem [{\citenamefont {St{\"o}hr}\ and\ \citenamefont
  {Siegmann}(2006)}]{stohrMagnetismFundamentalsNanoscale2006}%
  \BibitemOpen
  \bibfield  {author} {\bibinfo {author} {\bibfnamefont {J.}~\bibnamefont
  {St{\"o}hr}}\ and\ \bibinfo {author} {\bibfnamefont {H.~C.}\ \bibnamefont
  {Siegmann}},\ }\href@noop {} {\emph {\bibinfo {title} {Magnetism: {{From
  Fundamentals}} to {{Nanoscale Dynamics}}}}},\ Springer {{Series}} in
  {{Solid-State Sciences}}\ (\bibinfo  {publisher} {{Springer-Verlag}},\
  \bibinfo {address} {{Berlin Heidelberg}},\ \bibinfo {year}
  {2006})\BibitemShut {NoStop}%
\bibitem [{\citenamefont {White}(2007)}]{whiteQuantumTheoryMagnetism2007}%
  \BibitemOpen
  \bibfield  {author} {\bibinfo {author} {\bibfnamefont {R.~M.}\ \bibnamefont
  {White}},\ }\href@noop {} {\emph {\bibinfo {title} {Quantum Theory of
  Magnetism Magnetic Properties of Materials}}}\ (\bibinfo  {publisher}
  {{Springer}},\ \bibinfo {address} {{Berlin}},\ \bibinfo {year}
  {2007})\BibitemShut {NoStop}%
\bibitem [{\citenamefont
  {Goodenough}(1963)}]{goodenoughMagnetismChemicalBond1963}%
  \BibitemOpen
  \bibfield  {author} {\bibinfo {author} {\bibfnamefont {J.~B.}\ \bibnamefont
  {Goodenough}},\ }\href@noop {} {\emph {\bibinfo {title} {Magnetism {{And The
  Chemical Bond}}}}}\ (\bibinfo  {publisher} {{John Wiley And Sons}},\ \bibinfo
  {year} {1963})\BibitemShut {NoStop}%
\bibitem [{\citenamefont {Nguyen}\ and\ \citenamefont
  {Cava}(2021)}]{nguyenHexagonalPerovskitesQuantum2021}%
  \BibitemOpen
  \bibfield  {author} {\bibinfo {author} {\bibfnamefont {L.~T.}\ \bibnamefont
  {Nguyen}}\ and\ \bibinfo {author} {\bibfnamefont {R.~J.}\ \bibnamefont
  {Cava}},\ }\href {https://doi.org/10.1021/acs.chemrev.0c00622} {\bibfield
  {journal} {\bibinfo  {journal} {Chemical Reviews}\ }\textbf {\bibinfo
  {volume} {121}},\ \bibinfo {pages} {2935} (\bibinfo {year}
  {2021})}\BibitemShut {NoStop}%
\bibitem [{\citenamefont {Yin}\ \emph {et~al.}(2017)\citenamefont {Yin},
  \citenamefont {Tian}, \citenamefont {Li}, \citenamefont {Liao},\ and\
  \citenamefont {Lin}}]{yinNew10HPerovskites2017c}%
  \BibitemOpen
  \bibfield  {author} {\bibinfo {author} {\bibfnamefont {C.}~\bibnamefont
  {Yin}}, \bibinfo {author} {\bibfnamefont {G.}~\bibnamefont {Tian}}, \bibinfo
  {author} {\bibfnamefont {G.}~\bibnamefont {Li}}, \bibinfo {author}
  {\bibfnamefont {F.}~\bibnamefont {Liao}},\ and\ \bibinfo {author}
  {\bibfnamefont {J.}~\bibnamefont {Lin}},\ }\href
  {https://doi.org/10.1039/C7RA02183F} {\bibfield  {journal} {\bibinfo
  {journal} {RSC Advances}\ }\textbf {\bibinfo {volume} {7}},\ \bibinfo {pages}
  {33869} (\bibinfo {year} {2017})}\BibitemShut {NoStop}%
\bibitem [{\citenamefont {Nguyen}\ \emph {et~al.}(2019)\citenamefont {Nguyen},
  \citenamefont {Kong},\ and\ \citenamefont
  {Cava}}]{nguyenTrimersMnO6Octahedra2019c}%
  \BibitemOpen
  \bibfield  {author} {\bibinfo {author} {\bibfnamefont {L.~T.}\ \bibnamefont
  {Nguyen}}, \bibinfo {author} {\bibfnamefont {T.}~\bibnamefont {Kong}},\ and\
  \bibinfo {author} {\bibfnamefont {R.~J.}\ \bibnamefont {Cava}},\ }\href
  {https://doi.org/10.1088/2053-1591/ab0695} {\bibfield  {journal} {\bibinfo
  {journal} {Materials Research Express}\ }\textbf {\bibinfo {volume} {6}},\
  \bibinfo {pages} {056108} (\bibinfo {year} {2019})}\BibitemShut {NoStop}%
\bibitem [{\citenamefont {Moessner}\ and\ \citenamefont
  {Ramirez}(2006)}]{moessnerGeometricalFrustration2006}%
  \BibitemOpen
  \bibfield  {author} {\bibinfo {author} {\bibfnamefont {R.}~\bibnamefont
  {Moessner}}\ and\ \bibinfo {author} {\bibfnamefont {A.~P.}\ \bibnamefont
  {Ramirez}},\ }\href {https://doi.org/10.1063/1.2186278} {\bibfield  {journal}
  {\bibinfo  {journal} {Physics Today}\ }\textbf {\bibinfo {volume} {59}},\
  \bibinfo {pages} {24} (\bibinfo {year} {2006})}\BibitemShut {NoStop}%
\bibitem [{\citenamefont {Balents}(2010)}]{balentsSpinLiquidsFrustrated2010}%
  \BibitemOpen
  \bibfield  {author} {\bibinfo {author} {\bibfnamefont {L.}~\bibnamefont
  {Balents}},\ }\href {https://doi.org/10.1038/nature08917} {\bibfield
  {journal} {\bibinfo  {journal} {Nature}\ }\textbf {\bibinfo {volume} {464}},\
  \bibinfo {pages} {199} (\bibinfo {year} {2010})}\BibitemShut {NoStop}%
\bibitem [{\citenamefont {Zhou}\ \emph {et~al.}(2017)\citenamefont {Zhou},
  \citenamefont {Kanoda},\ and\ \citenamefont
  {Ng}}]{zhouQuantumSpinLiquid2017}%
  \BibitemOpen
  \bibfield  {author} {\bibinfo {author} {\bibfnamefont {Y.}~\bibnamefont
  {Zhou}}, \bibinfo {author} {\bibfnamefont {K.}~\bibnamefont {Kanoda}},\ and\
  \bibinfo {author} {\bibfnamefont {T.-K.}\ \bibnamefont {Ng}},\ }\href
  {https://doi.org/10.1103/RevModPhys.89.025003} {\bibfield  {journal}
  {\bibinfo  {journal} {Reviews of Modern Physics}\ }\textbf {\bibinfo {volume}
  {89}},\ \bibinfo {pages} {025003} (\bibinfo {year} {2017})}\BibitemShut
  {NoStop}%
\bibitem [{\citenamefont {Nikuni}\ \emph {et~al.}(2000)\citenamefont {Nikuni},
  \citenamefont {Oshikawa}, \citenamefont {Oosawa},\ and\ \citenamefont
  {Tanaka}}]{nikuniBoseEinsteinCondensationDilute2000}%
  \BibitemOpen
  \bibfield  {author} {\bibinfo {author} {\bibfnamefont {T.}~\bibnamefont
  {Nikuni}}, \bibinfo {author} {\bibfnamefont {M.}~\bibnamefont {Oshikawa}},
  \bibinfo {author} {\bibfnamefont {A.}~\bibnamefont {Oosawa}},\ and\ \bibinfo
  {author} {\bibfnamefont {H.}~\bibnamefont {Tanaka}},\ }\href
  {https://doi.org/10.1103/PhysRevLett.84.5868} {\bibfield  {journal} {\bibinfo
   {journal} {Physical Review Letters}\ }\textbf {\bibinfo {volume} {84}},\
  \bibinfo {pages} {5868} (\bibinfo {year} {2000})}\BibitemShut {NoStop}%
\bibitem [{\citenamefont {Aczel}\ \emph {et~al.}(2009)\citenamefont {Aczel},
  \citenamefont {Kohama}, \citenamefont {Marcenat}, \citenamefont {Weickert},
  \citenamefont {Jaime}, \citenamefont {{Ayala-Valenzuela}}, \citenamefont
  {McDonald}, \citenamefont {Selesnic}, \citenamefont {Dabkowska},\ and\
  \citenamefont {Luke}}]{aczelFieldInducedBoseEinsteinCondensation2009}%
  \BibitemOpen
  \bibfield  {author} {\bibinfo {author} {\bibfnamefont {A.~A.}\ \bibnamefont
  {Aczel}}, \bibinfo {author} {\bibfnamefont {Y.}~\bibnamefont {Kohama}},
  \bibinfo {author} {\bibfnamefont {C.}~\bibnamefont {Marcenat}}, \bibinfo
  {author} {\bibfnamefont {F.}~\bibnamefont {Weickert}}, \bibinfo {author}
  {\bibfnamefont {M.}~\bibnamefont {Jaime}}, \bibinfo {author} {\bibfnamefont
  {O.~E.}\ \bibnamefont {{Ayala-Valenzuela}}}, \bibinfo {author} {\bibfnamefont
  {R.~D.}\ \bibnamefont {McDonald}}, \bibinfo {author} {\bibfnamefont {S.~D.}\
  \bibnamefont {Selesnic}}, \bibinfo {author} {\bibfnamefont {H.~A.}\
  \bibnamefont {Dabkowska}},\ and\ \bibinfo {author} {\bibfnamefont {G.~M.}\
  \bibnamefont {Luke}},\ }\href
  {https://doi.org/10.1103/PhysRevLett.103.207203} {\bibfield  {journal}
  {\bibinfo  {journal} {Physical Review Letters}\ }\textbf {\bibinfo {volume}
  {103}},\ \bibinfo {pages} {207203} (\bibinfo {year} {2009})}\BibitemShut
  {NoStop}%
\bibitem [{\citenamefont {Deisenhofer}\ \emph {et~al.}(2006)\citenamefont
  {Deisenhofer}, \citenamefont {Eremina}, \citenamefont {Pimenov},
  \citenamefont {Gavrilova}, \citenamefont {Berger}, \citenamefont {Johnsson},
  \citenamefont {Lemmens}, \citenamefont {{Krug von Nidda}}, \citenamefont
  {Loidl}, \citenamefont {Lee},\ and\ \citenamefont
  {Whangbo}}]{deisenhoferStructuralMagneticDimers2006}%
  \BibitemOpen
  \bibfield  {author} {\bibinfo {author} {\bibfnamefont {J.}~\bibnamefont
  {Deisenhofer}}, \bibinfo {author} {\bibfnamefont {R.~M.}\ \bibnamefont
  {Eremina}}, \bibinfo {author} {\bibfnamefont {A.}~\bibnamefont {Pimenov}},
  \bibinfo {author} {\bibfnamefont {T.}~\bibnamefont {Gavrilova}}, \bibinfo
  {author} {\bibfnamefont {H.}~\bibnamefont {Berger}}, \bibinfo {author}
  {\bibfnamefont {M.}~\bibnamefont {Johnsson}}, \bibinfo {author}
  {\bibfnamefont {P.}~\bibnamefont {Lemmens}}, \bibinfo {author} {\bibfnamefont
  {H.-A.}\ \bibnamefont {{Krug von Nidda}}}, \bibinfo {author} {\bibfnamefont
  {A.}~\bibnamefont {Loidl}}, \bibinfo {author} {\bibfnamefont {K.-S.}\
  \bibnamefont {Lee}},\ and\ \bibinfo {author} {\bibfnamefont {M.-H.}\
  \bibnamefont {Whangbo}},\ }\href {https://doi.org/10.1103/PhysRevB.74.174421}
  {\bibfield  {journal} {\bibinfo  {journal} {Physical Review B}\ }\textbf
  {\bibinfo {volume} {74}},\ \bibinfo {pages} {174421} (\bibinfo {year}
  {2006})}\BibitemShut {NoStop}%
\bibitem [{\citenamefont {Singh}\ and\ \citenamefont
  {Johnston}(2007)}]{singhSingletGroundState2007}%
  \BibitemOpen
  \bibfield  {author} {\bibinfo {author} {\bibfnamefont {Y.}~\bibnamefont
  {Singh}}\ and\ \bibinfo {author} {\bibfnamefont {D.~C.}\ \bibnamefont
  {Johnston}},\ }\href {https://doi.org/10.1103/PhysRevB.76.012407} {\bibfield
  {journal} {\bibinfo  {journal} {Physical Review B}\ }\textbf {\bibinfo
  {volume} {76}},\ \bibinfo {pages} {012407} (\bibinfo {year}
  {2007})}\BibitemShut {NoStop}%
\bibitem [{\citenamefont {Uchida}\ \emph {et~al.}(2001)\citenamefont {Uchida},
  \citenamefont {Tanaka}, \citenamefont {Bartashevich},\ and\ \citenamefont
  {Goto}}]{uchidaSingletGroundState2001}%
  \BibitemOpen
  \bibfield  {author} {\bibinfo {author} {\bibfnamefont {M.}~\bibnamefont
  {Uchida}}, \bibinfo {author} {\bibfnamefont {H.}~\bibnamefont {Tanaka}},
  \bibinfo {author} {\bibfnamefont {M.~I.}\ \bibnamefont {Bartashevich}},\ and\
  \bibinfo {author} {\bibfnamefont {T.}~\bibnamefont {Goto}},\ }\href
  {https://doi.org/10.1143/JPSJ.70.1790} {\bibfield  {journal} {\bibinfo
  {journal} {Journal of the Physical Society of Japan}\ }\textbf {\bibinfo
  {volume} {70}},\ \bibinfo {pages} {1790} (\bibinfo {year}
  {2001})}\BibitemShut {NoStop}%
\bibitem [{\citenamefont {Uchida}\ \emph {et~al.}(2002)\citenamefont {Uchida},
  \citenamefont {Tanaka}, \citenamefont {Mitamura}, \citenamefont {Ishikawa},\
  and\ \citenamefont {Goto}}]{uchidaHighfieldMagnetizationProcess2002}%
  \BibitemOpen
  \bibfield  {author} {\bibinfo {author} {\bibfnamefont {M.}~\bibnamefont
  {Uchida}}, \bibinfo {author} {\bibfnamefont {H.}~\bibnamefont {Tanaka}},
  \bibinfo {author} {\bibfnamefont {H.}~\bibnamefont {Mitamura}}, \bibinfo
  {author} {\bibfnamefont {F.}~\bibnamefont {Ishikawa}},\ and\ \bibinfo
  {author} {\bibfnamefont {T.}~\bibnamefont {Goto}},\ }\href
  {https://doi.org/10.1103/PhysRevB.66.054429} {\bibfield  {journal} {\bibinfo
  {journal} {Physical Review B}\ }\textbf {\bibinfo {volume} {66}},\ \bibinfo
  {pages} {054429} (\bibinfo {year} {2002})}\BibitemShut {NoStop}%
\bibitem [{\citenamefont {Tsujii}\ \emph {et~al.}(2005)\citenamefont {Tsujii},
  \citenamefont {Andraka}, \citenamefont {Uchida}, \citenamefont {Tanaka},\
  and\ \citenamefont {Takano}}]{tsujiiSpecificHeatSpindimer2005}%
  \BibitemOpen
  \bibfield  {author} {\bibinfo {author} {\bibfnamefont {H.}~\bibnamefont
  {Tsujii}}, \bibinfo {author} {\bibfnamefont {B.}~\bibnamefont {Andraka}},
  \bibinfo {author} {\bibfnamefont {M.}~\bibnamefont {Uchida}}, \bibinfo
  {author} {\bibfnamefont {H.}~\bibnamefont {Tanaka}},\ and\ \bibinfo {author}
  {\bibfnamefont {Y.}~\bibnamefont {Takano}},\ }\href
  {https://doi.org/10.1103/PhysRevB.72.214434} {\bibfield  {journal} {\bibinfo
  {journal} {Physical Review B}\ }\textbf {\bibinfo {volume} {72}},\ \bibinfo
  {pages} {214434} (\bibinfo {year} {2005})}\BibitemShut {NoStop}%
\bibitem [{\citenamefont {Lai}\ and\ \citenamefont
  {Valldor}(2017)}]{laiCoexistenceSpinOrdering2017f}%
  \BibitemOpen
  \bibfield  {author} {\bibinfo {author} {\bibfnamefont {K.~T.}\ \bibnamefont
  {Lai}}\ and\ \bibinfo {author} {\bibfnamefont {M.}~\bibnamefont {Valldor}},\
  }\href {https://doi.org/10.1038/srep43767} {\bibfield  {journal} {\bibinfo
  {journal} {Scientific Reports}\ }\textbf {\bibinfo {volume} {7}},\ \bibinfo
  {pages} {43767} (\bibinfo {year} {2017})}\BibitemShut {NoStop}%
\bibitem [{\citenamefont {Garcia}\ \emph {et~al.}(2015)\citenamefont {Garcia},
  \citenamefont {Kaneko}, \citenamefont {Granado}, \citenamefont
  {Sichelschmidt}, \citenamefont {Holzel}, \citenamefont {Duque}, \citenamefont
  {Nunes}, \citenamefont {{Marques-Ferreira}},\ and\ \citenamefont
  {{Lora-Serrano}}}]{garciaMagneticDimersTrimers2015f}%
  \BibitemOpen
  \bibfield  {author} {\bibinfo {author} {\bibfnamefont {F.~A.}\ \bibnamefont
  {Garcia}}, \bibinfo {author} {\bibfnamefont {U.~F.}\ \bibnamefont {Kaneko}},
  \bibinfo {author} {\bibfnamefont {E.}~\bibnamefont {Granado}}, \bibinfo
  {author} {\bibfnamefont {J.}~\bibnamefont {Sichelschmidt}}, \bibinfo {author}
  {\bibfnamefont {M.}~\bibnamefont {Holzel}}, \bibinfo {author} {\bibfnamefont
  {J.~G.~S.}\ \bibnamefont {Duque}}, \bibinfo {author} {\bibfnamefont
  {C.~A.~J.}\ \bibnamefont {Nunes}}, \bibinfo {author} {\bibfnamefont
  {P.}~\bibnamefont {{Marques-Ferreira}}},\ and\ \bibinfo {author}
  {\bibfnamefont {R.}~\bibnamefont {{Lora-Serrano}}},\ }\href
  {https://doi.org/10.1103/PhysRevB.91.224416} {\bibfield  {journal} {\bibinfo
  {journal} {Physical Review B}\ }\textbf {\bibinfo {volume} {91}},\ \bibinfo
  {pages} {224416} (\bibinfo {year} {2015})}\BibitemShut {NoStop}%
\bibitem [{\citenamefont {Cantarino}\ \emph {et~al.}(2019)\citenamefont
  {Cantarino}, \citenamefont {Amaral}, \citenamefont {Freitas}, \citenamefont
  {Araujo}, \citenamefont {{Lora-Serrano}}, \citenamefont {Luetkens},
  \citenamefont {Baines}, \citenamefont {Brauninger}, \citenamefont {Grinenko},
  \citenamefont {Sarkar}, \citenamefont {Klauss}, \citenamefont {Andrade},\
  and\ \citenamefont {Garcia}}]{cantarinoDynamicMagnetismDisordered2019}%
  \BibitemOpen
  \bibfield  {author} {\bibinfo {author} {\bibfnamefont {M.~R.}\ \bibnamefont
  {Cantarino}}, \bibinfo {author} {\bibfnamefont {R.~P.}\ \bibnamefont
  {Amaral}}, \bibinfo {author} {\bibfnamefont {R.~S.}\ \bibnamefont {Freitas}},
  \bibinfo {author} {\bibfnamefont {J.~C.~R.}\ \bibnamefont {Araujo}}, \bibinfo
  {author} {\bibfnamefont {R.}~\bibnamefont {{Lora-Serrano}}}, \bibinfo
  {author} {\bibfnamefont {H.}~\bibnamefont {Luetkens}}, \bibinfo {author}
  {\bibfnamefont {C.}~\bibnamefont {Baines}}, \bibinfo {author} {\bibfnamefont
  {S.}~\bibnamefont {Brauninger}}, \bibinfo {author} {\bibfnamefont
  {V.}~\bibnamefont {Grinenko}}, \bibinfo {author} {\bibfnamefont
  {R.}~\bibnamefont {Sarkar}}, \bibinfo {author} {\bibfnamefont {H.~H.}\
  \bibnamefont {Klauss}}, \bibinfo {author} {\bibfnamefont {E.~C.}\
  \bibnamefont {Andrade}},\ and\ \bibinfo {author} {\bibfnamefont {F.~A.}\
  \bibnamefont {Garcia}},\ }\href {https://doi.org/10.1103/PhysRevB.99.054412}
  {\bibfield  {journal} {\bibinfo  {journal} {Physical Review B}\ }\textbf
  {\bibinfo {volume} {99}},\ \bibinfo {pages} {054412} (\bibinfo {year}
  {2019})}\BibitemShut {NoStop}%
\bibitem [{\citenamefont {Komleva}\ \emph {et~al.}(2020)\citenamefont
  {Komleva}, \citenamefont {Khomskii},\ and\ \citenamefont
  {Streltsov}}]{komlevaThreesiteTransitionmetalClusters2020}%
  \BibitemOpen
  \bibfield  {author} {\bibinfo {author} {\bibfnamefont {E.~V.}\ \bibnamefont
  {Komleva}}, \bibinfo {author} {\bibfnamefont {D.~I.}\ \bibnamefont
  {Khomskii}},\ and\ \bibinfo {author} {\bibfnamefont {S.~V.}\ \bibnamefont
  {Streltsov}},\ }\href {https://doi.org/10.1103/PhysRevB.102.174448}
  {\bibfield  {journal} {\bibinfo  {journal} {Physical Review B}\ }\textbf
  {\bibinfo {volume} {102}},\ \bibinfo {pages} {174448} (\bibinfo {year}
  {2020})}\BibitemShut {NoStop}%
\end{thebibliography}%

\section{Appendix}

\appendix

\section{Exact diagonalization}

\label{sec:appendixA} In the following development we present analytical
solutions to the problems of dimer and trimer clusters interactions,
and also to inter-cluster interactions. Our approach to study interactions
consists in find the eigenenergies of the respective Heisenberg Hamiltonian,
$\mathcal{H}=\sum_{ij}J_{ij}S_{i}\cdot S_{j}$, in which $S_{i}$
are spin operators and $J_{ij}$ are the exchange constants (usually
in units of $\text{K}$), related to the probability of electronic
hopping from site $i$ to site $j$. 

\subsection{Dimers}

\label{sec:dimorph}

Let us consider a dimer system consisting of two sites with electrons
which may hop to the neighbor and interact effectively with total
spins $S_{1}$ and $S_{2}$, and exchange constant $J_{1}$. The Hamiltonian
treating the interaction is the Heisenberg Hamiltonian showed in Equation
\ref{eq:hamdimer-1}.

\begin{equation}
\mathcal{H}_{\text{dim}}=J_{1}S_{1}\cdot S_{2}\label{eq:hamdimer-1}
\end{equation}

The basis which diagonalizes the operators $S_{1}^{2},S_{2}^{2},S_{d}^{2}=(S_{1}+S_{2})^{2}$,
and $S_{z}$ (of dimension $(2s_{1}+1)(2s_{2}+1)$) has eigenvectors
$|s_{d},m_{d}\rangle$, where $S_{d}^{2}|s_{d},m_{d}\rangle=\hbar s_{d}(s_{d}+1)|s_{d},m_{d}\rangle$,
and $S_{z}|s_{d},m_{d}\rangle=\hbar m_{d}|s_{d},m_{d}\rangle$. All
expected values in this subsection are taken with respect to these
eigenkets.

With this basis, it is possible to diagonalize the Hamiltonian in
Equation \ref{eq:hamdimer-1} for any values of spin operators $S_{1}$
and $S_{2}$, and find $E=\langle s_{d},m_{d}|\mathcal{H}|s_{d},m_{d}\rangle=J_{1}\langle S_{1}\cdot S_{2}\rangle$.
The trick here is to define a dimer quantum number such that $\langle S_{d}^{2}\rangle=\langle(S_{1}+S_{2})^{2}\rangle=\langle S_{1}^{2}+S_{2}^{2}-2S_{1}\cdot S_{2}\rangle=s_{d}(s_{d}+1)\implies\langle S_{1}\cdot S_{2}\rangle=\frac{1}{2}[s_{d}(s_{d}+1)-s_{1}(s_{1}+1)-s_{2}(s_{2}+1)]$.
The quantum number $s_{d}$ runs from $|s_{1}-s_{2}|$ through $s_{1}+s_{2}$
in steps of 1. Therefore, Equation \ref{eq:endim} holds for the eigenenergies
of the dimer system. 
\begin{equation}
E_{d}(s_{d})=\frac{J_{1}}{2}[s_{d}(s_{d}+1)-s_{1}(s_{1}+1)-s_{2}(s_{2}+1)].\label{eq:endim}
\end{equation}

Adding a Zeeman-type perturbation $\mathcal{H}_{Z}=-g\mu_{B}HS_{z}$
(in units of $\hbar$) in the z-direction can be done straightforwardly:
$E_{n}^{(1)}=-g\mu_{B}Hm_{d}$.

\subsection{Trimers}

\label{sec:trim} Differently of dimers, the susceptibility of trimers
does not goes to zero $T\rightarrow0$ , because each trimer will
still present a net magnetic moment. 

Now the Heisenberg Hamiltonian of a trimer cluster is written as in
Equation \ref{eq:hamtrimer}. Here, we specialize in the case $J_{12}=J_{23}=J_{1}$
and $J_{13}=J_{2}=\alpha J_{1}$, where $0\leq\alpha\leq1$, which
is adequate for systems of magnetic trimers forming as in the cases
depicted in figure \ref{fig:full}. Thus:
\begin{equation}
\mathcal{H}_{\text{trim}}=J_{1}(S_{1}\cdot S_{2}+S_{2}\cdot S_{3}+\alpha S_{1}\cdot S_{3})\label{eq:hamtrimer-1}
\end{equation}
To diagonalize $\mathcal{H}_{\text{trim}}$ is similar to the case
of the dimers. The basis containing good quantum numbers to the problem
will be the one which couples three angular momenta (of dimension
$(2s_{1}+1)(2s_{2}+1)(2s_{3}+1)$), and therefore diagonalizes simultaneously
the operators $S_{i}^{2},S_{d}^{2}=(S_{1}+S_{2})^{2},S^{2}=(S_{1}+S_{2}+S_{3})^{2}$,
and $S_{z}$. Therefore, the eigenstates are denoted $|s_{d},s,m\rangle$,
where $S_{d}^{2}|s_{d},s,m\rangle=\hbar s_{d}(s_{d}+1)|s_{d},s,m\rangle$,
$S^{2}|s_{d},s,m\rangle=\hbar s(s+1)|s_{d},s,m\rangle$, and $S_{z}|s_{d},s,m\rangle=\hbar m|s_{d},s,m\rangle$.
All expected values in this subsection are taken with respect to these
eigenkets.

As the objective is now obtain three dot products between spin operators
we first define the same dimer quantum number $s_{d}$ in order to
obtain $\langle S_{1}\cdot S_{2}\rangle$ and $\langle S_{2}\cdot S_{3}\rangle$.
Next we consider $S=S_{1}+S_{2}+S_{3}$ which results in a total spin
quantum number $\langle S^{2}\rangle=s(s+1)$. Working out $\langle S^{2}\rangle$
we isolate $\langle S_{1}\cdot S_{3}\rangle$ and use our definitions
for $S_{d}$. Finally with all dot products in hand we substitute
them in $\langle E\rangle$ and obtain Equation \ref{eq:entrim} for
the eigenenergies of the system: 

\begin{equation}
\begin{split}E_{t}(s,s_{d})=\frac{J}{2}\Big\{ s(s+1)-s_{3}(s_{3}+1)+s_{d}(s_{d}+1)(\alpha-1)\\
-\alpha\big[s_{1}(s_{1}+1)+s_{2}(s_{2}+1)\big]\Big\}
\end{split}
\label{eq:entrim}
\end{equation}

Now the quantum numbers $s_{d}$ and $s$ admit values as follows.
The dimer quantum number $s_{d}$ is a result of the coupling of two
spin angular momenta, with total spins $s_{1}$, and $s_{2}$, so
its minimum value is $|s_{1}-s_{2}|$, and its maximum is $s_{1}+s_{2}$,
varying in steps of $1$. In its turn, the quantum number $s$, and
its multiplicity $m$ are a result of the coupling between $S_{d}$,
and $S_{3}$, therefore the minimum of $s$ is $|s_{d}-s_{3}|$, while
its maximum is $s_{d}+s_{3}$, with the multiplicity $m$ running
from $m=-s$ to $m=s$. The trick here is to make the multiplicity
$m_{3}$ of the third spin vary for each value of $s_{d}$, giving
the total spin quantum number, $s=s_{d}+m_{3}$, within the range
of values $s$ is defined.

For example, take $s_{1}=s_{2}=s_{3}=1/2$, so $s_{d}=\{1,0\}$, while
the minimum of $s$ is $|s_{d}-s_{3}|=|1-1/2|=1/2$ and its maximum
is $s_{d}+s_{3}=1+1/2=3/2$; there are $(2s_{1}+1)(2s_{2}+1)(2s_{3}+1)=8$
states in total. In this way the possible energy states are $E_{t}(3/2,1)$
($s_{d}=1,m_{3}=1/2,s=s_{d}+m_{3}=3/2$), which is $4$-degenerated
because of $m$, $E_{t}(1/2,1)$ ($s_{d}=1,m_{3}=-1/2,s=s_{d}+m_{3}=1/2$),
$2$-degenerated because of $m$, and $E_{t}(1/2,0)$ ($s_{d}=0,m_{3}=1/2,s=s_{d}+m_{3}=3/2$),
which is $2$-degenerated because of $m$.

Adding a Zeeman-type perturbation $\mathcal{H}_{Z}=-g\mu_{B}HS_{z}$
(in units of $\hbar$) in the z-direction can be done straightforwardly:
$E_{n}^{(1)}=-g\mu_{B}Hm$.

\section{Cluster effective spin}

\label{sec:appendixB} To apply Equation $\ref{eq:Tc}$ to estimate
the critical temperature resulting from inter-cluster interactions,
we need to determine the effective spins of the clusters in our system.
Despite the system may be in several spin states running from $0$
through $s_{\text{tot}}$ (total sum of the cluster spins), excited
states do not take part in the low temperature magnetic response.
We thus adopt, in the case of trimers, as $s_{\text{eff}}$: we are
thus interested in the expected value of the $S_{z}$ operator with
respect to the cluster ground state.

Let us take the Heisenberg Hamiltonian of our cluster system $\mathcal{H}_{0}$
(Equations \ref{eq:hamdimer} or \ref{eq:hamtrimer}) and a Zeeman
term $\mathcal{H}'$ such that $\mathcal{H}=\mathcal{H}_{0}+\mathcal{H}'$.
We then numerically diagonalize it and determine the ground state
$|GS\rangle$, which in this case is a vector of length $2s_{\text{tot}}+1$.
In its turn, the $S_{z}$ operator is a $(2s_{\text{tot}}+1)\times(2s_{\text{tot}}+1)$
matrix so that the expected value of equation \ref{eq:seff} is just
a contraction of vectors with a matrix.

\begin{equation}
s_{\text{eff}}=\langle GS|S_{z}|GS\rangle.\label{eq:seff}
\end{equation}

\begin{figure}[H]
\centering{}\includegraphics[width=1\linewidth]{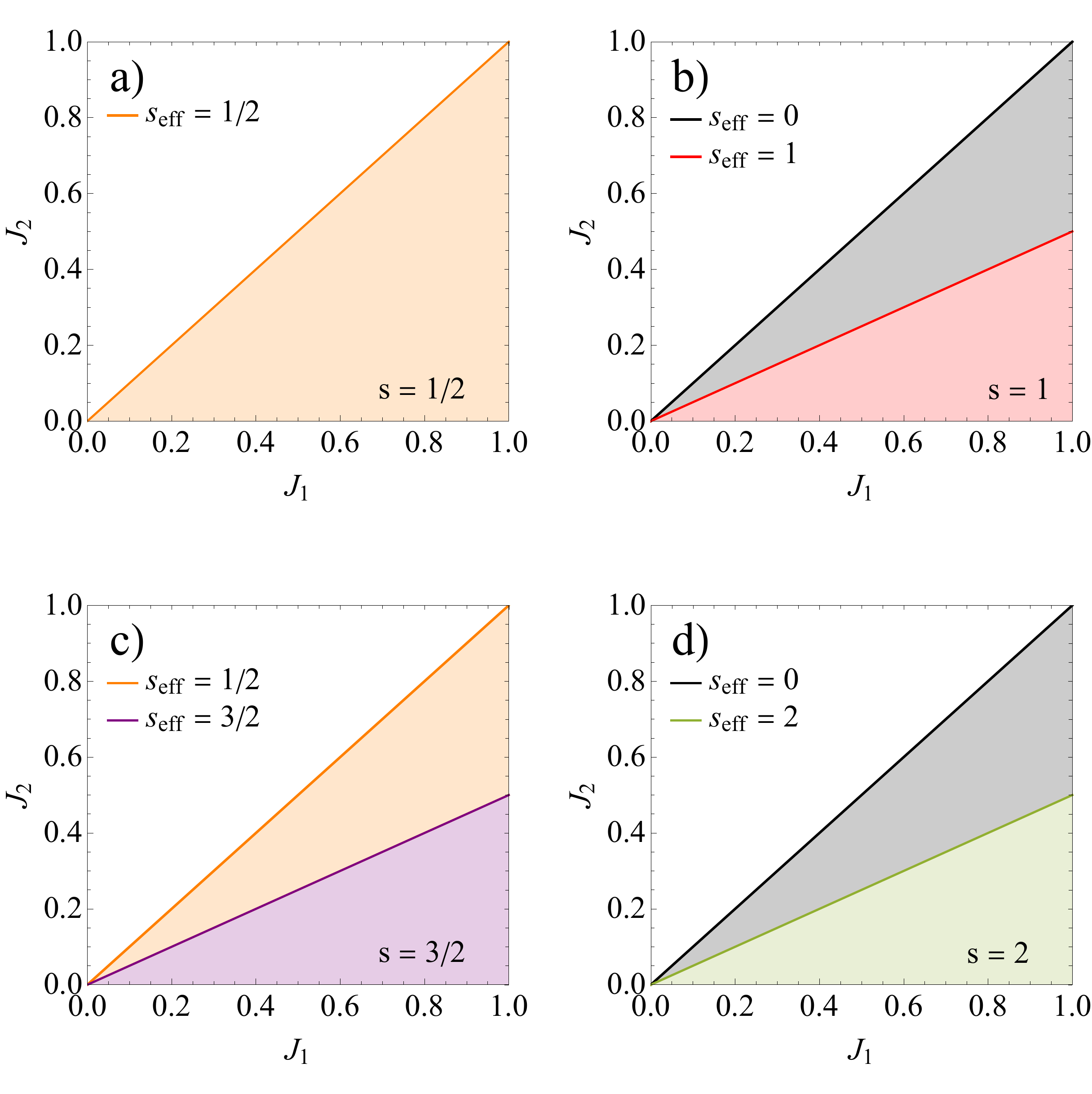}
\caption{$(a)-(d)$ The $J_{2}$ vs. $J_{1}$ trimers' ground state (colors)
phase diagrams of $s=1/2$, $s=1$, $s=3/2$, and $s=2$ spin trimers,
respectively . In all panels the upper straight lines denote $\alpha=J_{1}/J_{2}=1$,
while the bottom ones denote $\alpha=1/2$. Since in real systems
$\alpha<1$, white regions are unimportant. \label{fig:trimps}}
\end{figure}

Here we considered some usual cases for total spin as a function of
the relation between the first and, in the case of trimers, second
neighbors interactions. In Figure \ref{fig:trimps} we show the phase
space $J_{2}\times J_{1}$ for trimers of spins $s=1/2$, $s=1$,
$s=3/2$, and $s=2$. We identify a discrete behavior of the ground
state of the system as a function of the $\alpha=J_{2}/J_{1}$ relation.

\end{document}